\documentclass[referee]{cjaa}

\usepackage{graphicx}
\input{epsf.sty}
\input{psfig.sty}

\begin{document}
   \title{Afterglow Light Curves of Jetted Gamma-ray Burst Ejecta in Stellar Winds}
   %\volnopage{Vol.4 (2004) No.5, 455--472}      %%preserved for Editor. DOn't remove!
   \setcounter{page}{1}
   \author{Xue-Feng Wu
              \mailto{}
   \and Zi-Gao Dai
   \and Yong-Feng Huang
   \and Hai-Tao Ma
           }
   \institute{Department of Astronomy, Nanjing University, Nanjing 210093, China
             \email{xfwu@nju.edu.cn}}
   \offprints{X.-F. Wu}
   \date{Received~~2004 February 26; accepted~~2004~~May 9}

   \abstract{Optical and radio afterglows arising from the
   shocks by relativistic conical ejecta running into pre-burst
   massive stellar winds are revisited. Under the homogeneous thin-shell
   approximation and a realistic treatment for the lateral expansion of
   jets, our results show that a notable break exists in the optical light curve
   in most cases we calculated in which the physical parameters
   are varied within reasonable ranges. For a relatively tenuous wind which cannot
   decelerate the relativistic jet to cause a light curve break within days,
   the wind termination shock due to the ram pressure of the surrounding
   medium occurs at a small radius, namely, a few times $10^{17}$ cm.
   In such a structured wind environment, the jet will pass through the wind within several hours
   and run into the outer uniform dense medium. The resulting optical
   light curve flattens with a shallower drop after the jet encounters
   the uniform medium, and then declines deeply, triggered by runaway
   lateral expansion.
   }

   \authorrunning{X. F. Wu, Z. G. Dai, Y. F. Huang, \& H. T. Ma}          %author_head in even pages
   \titlerunning{Afterglow Light Curves from Jetted GRB Ejecta in Stellar Winds}
   \maketitle
%
%________________________________________________ sections below
%
\section{Introduction}
\label{sect:intro} Gamma-ray Bursts (GRBs) are becoming gradually
understood since the \textit{Beppo}SAX afterglow era (Piran 1999;
van Paradijs, Kouveliotou $\&$ Wijers 2000; Cheng $\&$ Lu 2001;
M\'{e}sz\'{a}ros 2002). Standard afterglow models have been set up
within the hydrodynamical context of relativistic external shocks
of spherical explosions running into either interstellar medium
(Sari, Piran $\&$ Narayan 1998) or stellar winds (Chevalier $\&$
Li 2000), with synchrotron emission as the main radiation
mechanism. Granot $\&$ Sari (2002) have obtained more accurate
results by using the well-known Blandford $\&$ McKee (1976)
self-similar inner structure of relativistic blastwave instead of
the thin shell approximation. However, the famous energy crisis
under isotropic assumption (Kulkarni et al. 1999), together with
the sharp decline of some well observed optical afterglows led to
the conjecture of a jet-like form of this phenomenon (Rhoads 1997;
Sari, Piran $\&$ Halpern 1999). Because of the relativistic
beaming effect, the early afterglow from a jet is no different
from a spherical blastwave. As the jet decelerates and the beaming
effect weakens, the whole surface of the jet unfolds to the
observer and the light curve declines more deeply with the
decreasing radiating extent compared to the spherical blastwave.
Additionally, the jet will decelerate much more rapidly when a
significant or runaway lateral expansion takes place. Based on the
homogeneous thin shell assumption, many authors have numerically
studied jet dynamics and the behavior of the afterglow light curve
(Panaitescu, M\'{e}sz\'{a}ros $\&$ Rees 1998; Huang et al. 2000a;
Huang, Dai $\&$ Lu 2000b; Moderski, Sikora $\&$ Bulik 2000).
Tremendous efforts have been devoted to the fitting of the GRB
afterglows of interest (Panaitescu $\&$ Kumar 2001, 2002). The jet
kinetic energy is found to be surprisingly similar in different
bursts, being tightly clustered around $3 \times 10^{50}$ erg.
Frail et al. (2001) found the genuine gamma-ray energy releases
are also clustered around $5\times 10^{50}$ erg after the jet
initial aperture $\theta_{\rm{j}}$ inferred from afterglows are
accounted for (a factor of 3 is amplified by Bloom, Frail $\&$
Kulkarni 2003). The total energy budget of a GRB is therefore
close to that of a common supernova. Late time X-ray luminosity
has been recognized to be largely independent of the types of
environment (Kumar 2000; Freedman $\&$ Waxman 2001). Subsequent
statistics of the X-ray luminosities of several GRBs at 10 hours
since bursts led to the conclusion that $L_{\rm{X, 10hr}}$
corrected by the jet aperture is again clustered around
$10^{44}-10^{45}$ erg s$^{-1}$ (Berger, Kulkarni $\&$ Frail
2003a). The above three independent estimations all identify the
GRB as a standard energy reservoir and hence as a possible probe
to the Universe.

The discovery of optical transient of GRB 030329 (Price et al. 2003) and its association with SN
2003dh (Stanek et al. 2003) provides strong evidence for GRB-Supernova connection and removes any
lingering doubt on the association of GRB 980425 with SN 1998bw. These two associations together
with several GRBs with late time re-brightening strongly imply near simultaneity of GRBs and SNe,
with trigger time difference no more than 1 to 2 days (Wu et al 2003; Hjorth et al. 2003). Although
the detailed physical processes are still uncertain, the central engine of at least long GRBs has
thus been confirmed to be the core collapse of massive stars, called collapsars by Woosley (1993)
as well as hypernovae for their energetics by Paczy\'{n}ski (1998). Previous studies favored the
interstellar medium (ISM) as the environment of GRBs, albeit there were indications of wind
environment for several GRBs within the spherical wind interaction model (Dai $\&$ Lu 1998b;
M\'{e}sz\'{a}ros, Rees $\&$ Wijers 1998; Chevalier $\&$ Li 1999, 2000; Li $\&$ Chevalier 1999,
2001, 2003; Dai $\&$ Wu 2003). The preference is partly due to the conclusion, based on the
previous works (Kumar $\&$ Panaitescu 2000; Gou et al. 2001), that a jet in a stellar wind cannot
produce a sharp break in the light curve. In fact, radiative loss of jet energy was improperly
neglected in these works, because the radiative phase lasts for hours or even one day due to the
large wind density in the early afterglow. This fast cooling radiation will reduce the jet energy
by about one order of magnitude. Realistic jets with energy losses in the stellar wind environments
are found to be consistent with several GRB afterglows (Panaitescu $\&$ Kumar 2002). Another origin
of the bias against the wind environment arises from the relatively large $\chi^2$ or even
impossibility of fit in some bursts. This result may be understood if the complicated mass loss
evolutions of the progenitors at different stages and their interactions with larger scale
environments are taken into account (Chevalier 2003). However, the region near the progenitor
(within sub-pc) will not be affected and will be reflected in the intra-day afterglow. Even in the
case of a supernova taking place 2 days earlier than the associated GRB, the supernova ejecta
moving with one tenth of the light speed will reach no more than $10^{16}$ cm, which is the typical
deceleration radius marking the beginning of the afterglows.

According to the above review of the recent progress of the GRBs and the wind environment, it seems
worthwhile to revisit the evolution and radiation of homogeneous jets in a stellar wind
environment. In this paper, we will study the dynamics and radiation in the jet plus wind model in
Section 2. Numerical results with afterglow light curves clarifying the effects of different
parameters are presented in Section 3. In Section 4 we give our discussion and conclusions.

\section{Dynamics and Radiation}
\label{sect:Dyn} The dynamics of a realistic jet with radiative loss and lateral expansion running
into a homogeneous interstellar medium is well described in Huang, Dai $\&$ Lu (2000b) and Huang et
al. (2000a). Gou et al. (2001) studied in detail the evolution of a jet in a stellar wind
environment, but they did not consider the radiative loss. We will follow these works and consider
the radiative loss which is especially important in the wind environment case. We will also extend
our calculation of the light curves into the radio band. Here we first give a brief description of
our improved model.

\subsection{Dynamics}
The evolution of the radius $R$, swept-up mass $m$, half-opening
angle $\theta$ and Lorentz factor $\gamma$ of the beamed GRB
ejecta with initial baryon loading $M_{\rm{ej}}$ and Lorentz
factor $\gamma_0$ is described by (Huang et al. 2000a; Huang, Dai
$\&$ Lu 2000b)
\begin{equation}
\frac{\rm{d}\it{R}}{\rm{d}\it{t}}=\beta c \gamma
(\gamma+\sqrt{\gamma^2-1}),
\end{equation}
\begin{equation}
\frac{\rm{d}\it{m}}{\rm{d}\it{R}}=2\pi
R^2(1-\cos{\theta})nm_{\rm{p}},
\end{equation}
\begin{equation}
\frac{\rm{d}\theta}{\rm{d}\it{t}}=\frac{c_{\rm{s}}(\gamma+\sqrt{\gamma^2-1})}{R},
\end{equation}
\begin{equation}
\frac{\rm{d}\gamma}{\rm{d}\it{m}}=-\frac{\gamma^2-1}{M_{\rm{ej}}+\epsilon
m+2(1-\epsilon)\gamma m},
\end{equation}
where $t$ is the observer's time,
$\beta=\sqrt{\gamma^2-1}/\gamma$, $n$ is the proton number density
of the surrounding medium, $m_p$ is the mass of a proton, and
$\epsilon$ is the radiative efficiency. For a stellar wind
environment, the number density is given by (Chevalier $\&$ Li
1999, 2000)
\begin{equation}
n=A R^{-2},
\end{equation}
where $A=\dot{M}/4\pi m_p v_w=3\times 10^{35}A_{\ast}$ cm$^{-1}$,
with $\dot{M}$ being the mass loss rate and
\begin{equation}
A_{\ast}=\frac{\dot{M}}{10^{-5}M_{\odot}\rm{yr}^{-1}}(\frac{v_w}{\rm{10^3km\;
s^{-1}}})^{-1}
\end{equation}
is the wind parameter. The comoving sound speed $c_{\rm{s}}$ is
(Huang et al. 2000),
\begin{equation}
c_{\rm{s}}^2=\hat{\gamma}(\hat{\gamma}-1)(\gamma-1)\frac{1}{1+\hat{\gamma}(\gamma-1)}c^2,
\end{equation}
where $\hat{\gamma}\approx(4\gamma+1)/(3\gamma)$ is the adiabatic
index, which is appropriate for both relativistic and
non-relativistic equation of state (Dai, Huang $\&$ Lu 1999).

To estimate the radiative efficiency $\epsilon$, we assume the shock-accelerated electrons and the
amplified magnetic field in the ejecta comoving frame carry a constant fraction, $\xi_{\rm{e}}$ and
$\xi_{\rm{B}}$, of the total thermal energy. The magnetic energy density and the minimum Lorentz
factor of the shock accelerated power law electrons in the comoving frame are then determined by
\begin{equation}
\frac{B^{\prime 2}}{8\pi}=\xi_{\rm{B}}
\frac{\hat{\gamma}\gamma+1}{\hat{\gamma}-1}(\gamma-1)nm_{\rm{p}}c^2,
\end{equation}
\begin{equation}
\gamma_{\rm{m}}=\xi_{\rm{e}}(\gamma-1)\frac{m_{\rm{p}}(p-2)}{m_{\rm{e}}(p-1)}+1,
\end{equation}
$m_{\rm{e}}$ being the electron mass and $p$, the index in the electron power law energy
distribution. The radiative efficiency of the ejecta is determined by a combination of the
available fraction of energy to radiation contained in the electrons and the efficiency of the
radiation mechanisms (Dai et al. 1999),
\begin{equation}
\epsilon=\xi_{\rm{e}} \frac{t_{\rm{syn}}^{\prime
-1}}{t_{\rm{syn}}^{\prime -1 }+t_{\rm{ex}}^{\prime -1}},
\end{equation}
where $t_{\rm{ex}}^{\prime}=R/(\gamma c)$ and $t_{\rm{syn}}^{\prime}=6\pi
m_{\rm{e}}c/(\sigma_{\rm{T}} B^{\prime 2} \gamma_{\rm{m}})$ are the comoving-frame expansion time
and synchrotron cooling time, respectively.

\subsection{Synchrotron radiation and self-absorption}
\label{sect:SSA} The distribution of electrons newly accelerated by the shock is assumed to be a
power law function of the electron kinetic energy. Recently Huang $\&$ Cheng (2003) stressed that
the distribution function should take the following form,
$dN_{\rm{e}}^{\prime}/d\gamma_{\rm{e}}\propto (\gamma_{\rm{e}}-1)^{-p}$ with
$\gamma_{\rm{m}}\leq\gamma_{\rm{e}}\leq\gamma_{\rm{M}}$, where
$\gamma_{\rm{M}}=10^8(B^{\prime}/1\rm{G})^{-1/2}$. This is especially important in the deep
Newtonian stage. Radiation will cool down the electrons and thus change the shape of the
distribution. The cooling effect is significant for electrons with Lorentz factors above the
critical value (Sari, Piran $\&$ Narayan 1998),
\begin{equation}
\gamma_{\rm{c}}=\frac{6\pi m_{\rm{e}} c}{\sigma_{\rm{T}} \gamma
B^{\prime 2} t}.
\end{equation}

In the comoving frame, the synchrotron radiation power at
frequency $\nu^{\prime}$ from electrons of known distribution is
given by (Rybicki $\&$ Lightman 1979)
\begin{equation}
P^{\prime}(\nu^{\prime})=\frac{\sqrt{3} q_{\rm{e}}^3
B^{\prime}}{m_{\rm{e}}
c^2}\int_{\rm{min}(\gamma_{\rm{m}},\gamma_{\rm{c}})}^{\gamma_{\rm{M}}}(\frac{\rm{d}\it{N}_{\rm{e}}^{\prime}}{\rm{d}\gamma_{\rm{e}}})F(\frac{\nu^{\prime}}{\nu_{\rm{e}}^{\prime}})\rm{d}\gamma_{\rm{e}},
\end{equation}
where $q_{\rm{e}}$ is the electron charge, $\nu_{\rm{e}}^{\prime}=3\sin{\vartheta}\gamma_{\rm{e}}^2
q_{\rm{e}}B^{\prime}/4\pi m_{\rm{e}} c$ is the typical emission frequency of the $\gamma_{\rm{e}}$
electron, and
\begin{equation}
F(x)=x\int_{x}^{\infty}K_{5/3}(k)\rm{d}\it{k},
\end{equation}
with $K_{5/3}(k)$ the Bessel function. As emphasized by Wijers $\&$ Galama (1999), the random pitch
angle $\vartheta$ between the velocity of the electron and the magnetic field will have some
effects on the modelling of GRB afterglows. We choose the isotropic distribution of $\vartheta$ and
have
\begin{equation}
\nu_{\rm{e}}^{\prime}=\frac{3\gamma_{\rm{e}}^2 q_{\rm{e}}
B^{\prime}}{16m_{\rm{e}} c}.
\end{equation}
The characteristic frequencies corresponding to $\gamma_{\rm{c}}$, $\gamma_{\rm{m}}$ and
$\gamma_{\rm{M}}$ electrons are denoted by $\nu_{\rm{c}}^{\prime}$, $\nu_{\rm{m}}^{\prime}$ and
$\nu_{\rm{M}}^{\prime}$.

In the wind environment the early radio afterglow flux density is reduced significantly by
synchrotron self absorption (SSA). This effect can be calculated by the analytical expressions
derived by Wu et al. (2003). We give the more direct and convenient formula for the optical depth
by SSA in different electron distributions as follows:

1. For $1\leq\gamma_{\rm{c}}\leq\gamma_{\rm{m}}$,

\parbox{10cm}
{\[\displaystyle
\frac{\rm{d}\it{N}_{\rm{e}}^{\prime}}{\rm{d}\gamma_{\rm{e}}}=\left
\{
\begin{array}{ll}
C_0(\gamma_{\rm{e}}-1)^{-2}            & (\gamma_{\rm{c}}\leq\gamma_{\rm{e}}<\gamma_{\rm{m}}),  \\
C_1(\gamma_{\rm{e}}-1)^{-(p+1)}        &
(\gamma_{\rm{m}}\leq\gamma_{\rm{e}}\leq\gamma_{\rm{M}}),
\end{array} \right.
 \]} \hfill \parbox[b][0.5cm][b]{1cm}{\begin{equation}\end{equation}}
\begin{flushleft}where
\begin{equation}
C_0=[(\frac{1}{\gamma_{\rm{c}}-1}-\frac{1}{\gamma_{\rm{m}}-1})+\frac{1}{p(\gamma_{\rm{m}}-1)}(1-\frac{(\gamma_{\rm{m}}-1)^p}
{(\gamma_{\rm{M}}-1)^p})]^{-1}N_{\rm{ele}},
\end{equation}
\begin{equation}
C_1=C_0(\gamma_{\rm{m}}-1)^{p-1},
\end{equation}
and $N_{\rm{ele}}$ is the total electron number of a jet element.
The self-absorption optical depth is \end{flushleft}

\parbox{10cm}
{\[\displaystyle
\tau_{\nu^{\prime}}=c_{\rm{sa}}\frac{q_{\rm{e}}}{B^{\prime}}C_0\frac{N_{\rm{col}}}{N_{\rm{ele}}}\times\left
\{
\begin{array}{ll}
\gamma_{\rm{c}}^{-6}(\displaystyle\frac{\nu^{\prime}}{\nu_{\rm{c}}^{\prime}})^{-5/3}            & (\nu^{\prime}\leq\nu_{\rm{c}}^{\prime}),  \\
\gamma_{\rm{c}}^{-6}(\displaystyle\frac{\nu^{\prime}}{\nu_{\rm{c}}^{\prime}})^{-3}              & (\nu_{\rm{c}}^{\prime}<\nu^{\prime}\leq\nu_{\rm{m}}^{\prime}),  \\
\gamma_{\rm{m}}^{-6}(\displaystyle\frac{\nu^{\prime}}{\nu_{\rm{m}}^{\prime}})^{-(p+5)/2}  & (\nu_{\rm{m}}^{\prime}<\nu^{\prime}\leq\nu_{\rm{M}}^{\prime}),  \\
\gamma_{\rm{m}}^{-6}(\displaystyle\frac{\nu^{\prime}}{\nu_{\rm{m}}^{\prime}})^{-5/2}(\frac{\gamma_{\rm{m}}}{\gamma_{\rm{M}}})^{p}e^{1-\nu^{\prime}/\nu_{\rm{M}}^{\prime}}
& (\nu_{\rm{M}}^{\prime}<\nu^{\prime}),
\end{array} \right.
 \]} \hfill \parbox[b][0.5cm][b]{1cm}{\begin{equation}\end{equation}}
\begin{flushleft}where $c_{\rm{sa}}$ is (Wu et al. 2003)
\begin{equation}
c_{\rm{sa}}=10.4(p+2)/(p+2/3),
\end{equation}
while
\begin{equation}
N_{\rm{col}}=\frac{m}{2\pi(1-\cos{\theta})R^2 m_{\rm{p}}}
\end{equation}
is the column density through which the synchrotron photons will experience synchrotron self
absorption before emerging from the jet surface. We do not consider here the corresponding
correction on the optical depth by the exact distribution of electrons, since the SSA coefficients
are deduced under the assumption of soft photons and ultra-relativistic electrons, and when the
bulk of the electrons are in the non-relativistic region the emitting source has already been
optically thin to SSA.\end{flushleft}

2. For $\gamma_{\rm{m}}<\gamma_{\rm{c}}\leq\gamma_{\rm{M}}$,

\parbox{10cm}
{\[\displaystyle
\frac{\rm{d}\it{N}_{\rm{e}}^{\prime}}{\rm{d}\gamma_{\rm{e}}}=\left
\{
\begin{array}{ll}
C_2(\gamma_{\rm{e}}-1)^{-p}            & (\gamma_{\rm{m}}\leq\gamma_{\rm{e}}<\gamma_{\rm{c}}),  \\
C_3(\gamma_{\rm{e}}-1)^{-(p+1)}        &
(\gamma_{\rm{c}}\leq\gamma_{\rm{e}}\leq\gamma_{\rm{M}}),
\end{array} \right.
 \]} \hfill \parbox[b][0.5cm][b]{1cm}{\begin{equation}\end{equation}}
\begin{flushleft}where (Huang $\&$ Cheng 2003)
\begin{equation}
C_2=C_3/(\gamma_{\rm{c}}-1),
\end{equation}
\begin{equation}
C_3=[\frac{(\gamma_{\rm{m}}-1)^{1-p}-(\gamma_{\rm{c}}-1)^{1-p}}{(\gamma_{\rm{c}}-1)(p-1)}+\frac{(\gamma_{\rm{c}}-1)^{-p}-(\gamma_{\rm{M}}-1)^{-p}}{p}]^{-1}N_{\rm{ele}}.
\end{equation}
\end{flushleft}
\begin{flushleft}The optical depth is \end{flushleft}

\parbox{10cm}
{\[\displaystyle
\tau_{\nu^{\prime}}=c_{\rm{sa}}\frac{q_{\rm{e}}}{B^{\prime}}C_2\frac{N_{\rm{col}}}{N_{\rm{ele}}}\times\left
\{
\begin{array}{ll}
\gamma_{\rm{m}}^{-(p+4)}(\displaystyle\frac{\nu^{\prime}}{\nu_{\rm{m}}^{\prime}})^{-5/3}        & (\nu^{\prime}\leq\nu_{\rm{m}}^{\prime}),  \\
\gamma_{\rm{m}}^{-(p+4)}(\displaystyle\frac{\nu^{\prime}}{\nu_{\rm{m}}^{\prime}})^{-(p+4)/2}    & (\nu_{\rm{m}}^{\prime}<\nu^{\prime}\leq\nu_{\rm{c}}^{\prime}),  \\
\gamma_{\rm{c}}^{-(p+4)}(\displaystyle\frac{\nu^{\prime}}{\nu_{\rm{c}}^{\prime}})^{-(p+5)/2}          & (\nu_{\rm{c}}^{\prime}<\nu^{\prime}\leq\nu_{\rm{M}}^{\prime}),  \\
\gamma_{\rm{c}}^{-(p+4)}(\displaystyle\frac{\nu^{\prime}}{\nu_{\rm{c}}^{\prime}})^{-5/2}(\frac{\gamma_{\rm{c}}}{\gamma_{\rm{M}}})^{p}e^{1-\nu^{\prime}/\nu_{\rm{M}}^{\prime}}
& (\nu_{\rm{M}}^{\prime}<\nu^{\prime}).
\end{array} \right.
 \]} \hfill \parbox[b][0.5cm][b]{1cm}{\begin{equation}\end{equation}}

3. For $\gamma_{\rm{c}}>\gamma_{\rm{M}}$, we have

\begin{equation}
\frac{\rm{d}\it{N}_{\rm{e}}^{\prime}}{\rm{d}\gamma_{\rm{e}}}=C_4(\gamma_{\rm{e}}-1)^{-p}
\quad (\gamma_{\rm{m}}\leq\gamma_{\rm{e}}\leq\gamma_{\rm{M}}),
\end{equation}
where (Huang $\&$ Cheng 2003)
\begin{equation}
C_4=\frac{p-1}{(\gamma_{\rm{m}}-1)^{1-p}-(\gamma_{\rm{M}}-1)^{1-p}}N_{\rm{ele}}.
\end{equation}
The optical depth is

\parbox{10cm}
{\[\displaystyle
\tau_{\nu^{\prime}}=c_{\rm{sa}}\frac{q_{\rm{e}}}{B^{\prime}}C_4\frac{N_{\rm{col}}}{N_{\rm{ele}}}\times\left
\{
\begin{array}{ll}
\gamma_{\rm{m}}^{-(p+4)}(\displaystyle\frac{\nu^{\prime}}{\nu_{\rm{m}}^{\prime}})^{-5/3}        & (\nu^{\prime}\leq\nu_{\rm{m}}^{\prime}),  \\
\gamma_{\rm{m}}^{-(p+4)}(\displaystyle\frac{\nu^{\prime}}{\nu_{\rm{m}}^{\prime}})^{-(p+4)/2}    & (\nu_{\rm{m}}^{\prime}<\nu^{\prime}\leq\nu_{\rm{M}}^{\prime}),  \\
\gamma_{\rm{m}}^{-(p+4)}(\displaystyle\frac{\nu^{\prime}}{\nu_{\rm{m}}^{\prime}})^{-5/2}(\frac{\gamma_{\rm{m}}}{\gamma_{\rm{M}}})^{p-1}e^{1-\nu^{\prime}/\nu_{\rm{M}}^{\prime}}
& (\nu_{\rm{M}}^{\prime}<\nu^{\prime}).
\end{array} \right.
 \]} \hfill \parbox[b][0.5cm][b]{1cm}{\begin{equation}\end{equation}}

The radiation power is assumed to be isotropic in the comoving frame. Let $\Theta$ be the angle
between the velocity of an emitting element and the line of sight and define $\mu=\cos{\Theta}$,
the observed flux density at frequency $\nu$ from this emitting element is
\begin{equation}
S_{\nu}=\frac{1+z}{\gamma^3(1-\beta\mu)^3}\frac{1}{4\pi
D_{\rm{L}}^2(z)}\frac{1-\rm{exp}(-\tau[\gamma(1-\beta\mu)(1+z)\nu])}{\tau[\gamma(1-\beta\mu)(1+z)\nu]}P^{\prime}[\gamma(1-\beta\mu)(1+z)\nu],
\end{equation}
where the luminosity distance is
\begin{equation}
D_{\rm{L}}(z)=(1+z)\frac{c}{H_0}\int_{0}^{z}\frac{dz^{\prime}}{\sqrt{\Omega_{\rm{M}}
(1+z^{\prime})^3+\Omega_{\Lambda}}},
\end{equation}
with $\Omega_{\rm{M}}=0.3$, $\Omega_{\Lambda}=0.7$ and $H_0=65$
km/s/Mpc adopted in our calculations. The total observed flux
density can be integrated over the equal arrival time surface
determined by
\begin{equation}
t_{\rm{obs}}=(1+z)t=(1+z)\int\frac{1-\beta\mu}{\beta
c}\rm{d}\it{R}\equiv \rm{const}.
\end{equation}

\section{NUMERICAL RESULTS}
\label{res} We investigate the effects of various parameters on the afterglow light curve in our
realistic jet$+$wind model. For convenience, we chose a set of `standard' initial parameters as
follows: $E_{\rm{iso}}=2\times 10^{53}$ erg, $\theta_0=0.1$, $\theta_{\rm{v}}=0$, $\gamma_0=300$,
$\xi_{\rm{e}}=0.1$, $\xi_{\rm{B}}=0.1$, $p=2.2$, and $A_{\ast}=1.0$. $\theta_0$ is the jet
half-opening angle while $\theta_{\rm{v}}$ is the observer's viewing angle with respect to the jet
axis. The GRB in our calculations is assumed to be at $z=1$, and the luminosity distance is about
$D_{\rm{L}}=7.1$ Gpc.

Figure 1 shows the optical and radio light curves with $E_{\rm{iso}}$ varying between $10^{52}$ and
$10^{54}$ erg while the other parameters are fixed at their standard values. The early optical
light curves resemble roughly those of spherical ones (for analytical jet+wind model, see Livio
$\&$ Waxman 2000). The theoretical temporal evolution of the optical flux density changes from
$t^{-1/4}$ to $t^{-(3p-2)/4}$ ($~ t^{-1.15}$ for $p=2.2$) when $\nu_{\rm{m}}$ crosses the optical
band, at time (Chevalier $\&$ Li 2000)
\begin{equation}
t_{\rm{m}}=0.1(\frac{1+z}{2})^{1/3}(\frac{\xi_{\rm{e}}}{0.1})^{4/3}(\frac{\xi_{\rm{B}}}{0.1})^{1/3}
E_{\rm{iso,53}}^{1/3}\nu_{\rm{R}}^{-2/3}\;\rm{days},
\end{equation}
where $\nu_{\rm{R}}=\nu/4.36\times 10^{14}\;\rm{Hz}$ is scaled to the $R-$band frequency and where
we have adopted the convention $Q=10^{\rm{x}}Q_{\rm{x}}$. The actual indices of the light curves in
our calculations deviate slightly from the ideal spherical predictions. Wei $\&$ Lu (2002a, 2002b)
suggested that some sharp breaks come from a spectral origin. Nevertheless, the radiative loss of
kinetic energy and the effect of equal arrival time surface (EATS) cause several differences in the
temporal evolution. For the extremely high isotropic energy case, viz. $E_{\rm{iso,53}}=10$, there
are two breaks in the light curve, as it evolves from $t^{-0.95}$ to $t^{-1.6}$, and finally
approaches the jet-like behavior $t^{-2.0}$. This temporal behavior is also seen in the `standard'
$E_{\rm{iso,53}}=2$ case, except that the scaling law evolves from $t^{-1.14}$ to $t^{-1.8}$ and
finally reaches $t^{-2.0}$. For other low $E_{\rm{iso}}$ cases, the light curves have only one
break, while before the jet break the temporal index $\beta$ is $\sim 1.3 - 1.4$ and after the jet
break $\beta\approx 2.0$ ($F_{\nu}\propto t^{-\beta}$). The theoretical change of $\beta$ in the
spherical wind model takes place at (see also, Chevalier $\&$ Li 2000)
\begin{equation}
t_{\rm{c}}=3.8\times
10^3(\frac{1+z}{2})^{3}(\frac{\xi_{\rm{B}}}{0.1})^{3}E_{\rm{iso,53}}^{-1}A_{\ast}^{4}\nu_{\rm{R}}^2
\;\rm{days},
\end{equation}
which is much later than the moment when significant lateral expansion is included. After that
$\nu_{\rm{c}}$ will cross the $R$ band and $\beta$ changes from $(3p-2)/4$ to $(3p-1)/4$ (i.e. from
$1.15$ to $1.4$ for $p=2.2$). The difference of $\beta$ between the jet model and the spherical
model is ascribed to the EATS effect (see also Fig. 2 of Kumar $\&$ Panaitescu 2000). The radiative
loss of energy also plays an important role for this diversity of the light curves. A crude
estimate of the energy loss can be made by synchrotron radiation within the spherical fireball
model. When the fireball is in the fast cooling stage, the observed power is
$P_{\oplus}=-\displaystyle{\frac{\rm{d}\it{E_{\rm{iso}}}}{\rm{d}\it{t_{\oplus}}}\approx\xi_{\rm{e}}\frac{E_{\rm{iso}}}{t_{\oplus}}}$
and the fireball energy evolves as
\begin{equation}
E_{\rm{iso}}=E_{\rm{iso,i}}(\frac{t_{\oplus}}{t_{\rm{i}}})^{-\xi_{\rm{e}}},
\end{equation}
where $E_{\rm{iso,i}}$ is the initial energy and
$t_{\rm{i}}=t_{\rm{dec}}=1.5\times10^{-6}\displaystyle{(\frac{1+z}{2})E_{\rm{iso,53}}A_{\ast}^{-1}(\frac{\gamma_0}{300})^{-4}}\;\rm{day}$
is the initial time of the afterglow, which begins at the deceleration radius. The fast cooling
stage ends at
$t_0=2\displaystyle{(\frac{1+z}{2})(\frac{\xi_{\rm{e}}}{0.1})(\frac{\xi_{\rm{B}}}{0.1})A_{\ast}}
\;\rm{days}$ (Chevalier $\&$ Li 2000). The kinetic energy decreases to $E_{\rm{iso,0}}\sim 0.25
E_{\rm{iso,i}}$ ($\xi_{\rm{e}}=0.1$) at $t_0$. The synchrotron power in the slow cooling stage is
$P_{\oplus}\approx
\displaystyle{\frac{\xi_{\rm{e}}}{3-p}(\frac{t_{\oplus}}{t_0})^{-(p-2)/2}\frac{E_{\rm{iso}}}{t_{\oplus}}}$.
The fireball energy is
\begin{equation}
E_{\rm{iso,f}}\approx
E_{\rm{iso,0}}\rm{exp}[-\displaystyle\frac{2\xi_{\rm{e}}}{(3-p)(p-2)}]\approx
0.07\it{E_{\rm{iso,i}}},
\end{equation}
in which we have put $p=2.2$ and $\xi_{\rm{e}}=0.1$. Early works neglected such tremendous loss of
the jet energy and assumed relatively large isotropic energy and jet opening angle, leading to
inconspicuous jet breaks (Gou et al. 2001). However, we show that notable breaks of optical light
curves exist in all cases, as can be seen in Fig. 1. Although the temporal indices both before and
after the break (see Fig. 1, $t\sim$several$\times 10^4$ seconds) vary with $E_{\rm{iso}}$, the
change of the temporal index in each case is nearly the same, $\Delta\beta\approx 0.65$, and is
well within one decade of time. Kumar $\&$ Panaitescu (2000) found that in a stellar wind $\beta$
increases by $\sim 0.4$ over two decades of time, while in a uniform density medium $\beta$
increases by $\sim 0.7$ within about one decade of time. However, their conclusion is based on
ignoring the energy losses by radiation. Our present results thus make the jet+wind model
competitive with the jet+ISM model. Radio afterglows are affected by the synchrotron
self-absorption due to the dense wind medium at early times. A rapid rise of radio flux density
proportional to $t^{2}$ in Fig. 1 is consistent with theoretical expectation. As $E_{\rm{iso}}$
decreases, the transition to  the optical thin regime becomes later than $t_{\rm{m}}$, and the
radio light curve changes from type D to type E of Chevalier $\&$ Li (2000). At late times, the
radio afterglows decay as $t^{-p}$ because of lateral expansion.

The effect of the wind parameter $A_{\ast}$ on the afterglow light curves is shown in Fig. 2. The
early optical light curves (less than $\sim10^{3}$ s) and late-time radio light curves (when jet
lateral expansion is significant) show little difference in the three cases with different
$A_{\ast}$. The jet break in the optical band is indistinctive when $A_{\ast}=0.3$, contrary to
denser wind cases, in which an obvious optical break occurs around one day. The early radio
afterglow is suppressed more strongly in denser winds. The radio light curves changes from type D
to type E as $A_{\ast}$ increases.

For a relatively tenuous wind which cannot decelerate the relativistic jet to cause a sharp break
around one day, the wind termination shock due to the ram pressure balanced by the surrounding
medium occurs at a small radius, i.e. several times $10^{17}$ cm (Ramirez-Ruiz et al. 2001).
Different mass loss histories of the progenitors or different large-scale environments surrounding
the winds will lead to diverse environments for the GRB afterglow (see the review by Chevalier
2003). At least four kinds of GRB environment have been considered so far, stellar wind (Dai $\&$
Lu 1998b; M\'{e}sz\'{a}ros et al. 1998; Chevalier $\&$ Li 1999), the ISM, dense medium (Dai $\&$ Lu
1999) and density jumps (Dai $\&$ Lu 2002). The complex, stratified medium resulting from
interaction between the winds or between a wind and an outer dense medium interaction has been
recently proposed to explain GRB $030226$ by Dai $\&$ Wu (2003). This complicated and more
realistic environment has the potential of unifying the diverse media, and, especially, of
explaining the peculiar afterglow of GRB $030329$ that included a large flux increase and several
fluctuations. Fluctuations subtracted from the power-law light curve of GRB $021004$ indicate
clouds and shells around the wind of a Wolf-Rayet star, which is the progenitor of an SN Ib/c and a
collapsar (Schaefer et al. 2003; Mirabal et al. 2003). In the complicated wind case, the jet will
reach the wind termination shock radius at the observed time (Ramirez-Ruiz et al. 2001; Chevalier
$\&$ Li 2000)
\begin{equation}
t_{\rm{t}}=1.5(\frac{1+z}{2})A_{\ast,-1}^2
E_{\rm{iso,53}}^{-1}n_{\rm{out}} \;\rm{hrs},
\end{equation}
where $n_{\rm{out}}$ is the number density of the outer uniform medium in units of cm$^{-3}$. After
this time the jet enters the outer uniform medium. We calculate the light curve of such a jet with
$A_{\ast}=0.1$ and $n_{\rm{out}}=1$ cm$^{-3}$. The density ratio of the outer medium to the wind at
the termination shock radius is $\sim 4$, which will not lead to a reverse shock propagating into
the jet. As illustrated in Fig. 3, the resulting optical light curve flattens from an initial
$t^{-1.5}$ before 6 hours to $t^{-1.2}$ after entering the uniform medium, and then declines
steeply as $t^{-2.3}$ after $\sim$ 9 days due to runaway lateral expansion. This result provides
the first piece of evidence that in the tenuous wind case an obvious sharp break can be caused by
the medium outside the wind termination shock. The radio flux density will increase by a factor of
a few since $t_{\rm{t}}$ and thereafter follows the behavior of the jet+ISM model, which exhibits a
jet break and shows a late time flattening when the shock becomes spherical and enters the deep
Newtonian phase.

The previous estimate of the jet half-opening angle $\theta_0$ comes from direct fitting of
afterglows and observed jet break time (Panaitescu $\&$ Kumar 2001, 2002; Frail et al. 2001; Bloom
et al. 2003). Statistics shows a large dispersion of $\theta_0$, ranging from $2^{\circ}$ to $\sim
40^{\circ}$. Here we calculate the effect of $\theta_0$ on the light curves, this result is shown
in Fig. 4. A large $\theta_0$ reduces the jet edge effects. For the case of $\theta_0=0.2$, the
temporal index $\beta$ for the $R$ band varies from $1.1$ to $1.55$ and then to $2.0$ at $\sim$
$100$ days. However, jets with relatively smaller $\theta_0$ will experience the jet break at
earlier times. For the cases of $\theta_0=0.075$ and $0.05$, $\beta$ evolves from $1.3-1.4$, to
$2.0$ at around 1 day. The corresponding $\Delta\beta$ is $0.7$ and $0.6$, respectively. The radio
light curves show no difference at early times, due to the relativistic beaming effect and the same
energy per solid angle.

As we know, the strong correlation between $E_{\rm{iso}}$ and $\theta_0$ makes the GRBs a standard
candle. A reliable treatment of the effect of $E_{\rm{iso}}$ or $\theta_0$ to the jet light curves
should include their intrinsic connection. We calculate the afterglows under the assumption of
standard energy reservoir, $E_{\rm{j}}\approx
E_{\rm{iso}}\displaystyle\frac{\theta_0^2}{2}\equiv2\times10^{51}$ erg, and illustrate the results
in Fig. 5. The characteristic feature of sharp breaks still remains for the narrow jets. For the
case of $\theta_0=0.2$, $\beta$ for the $R$ band evolves from $1.16$ to $1.7$, which is relative
shallower than typical jet-break. The temporal index will further approach $1.86$ tens of days
later. For a standard candle with smaller angle of $\theta_0=0.05$ ($\theta_0=0.025$), $\beta$
evolves from $1.25$ ($1.3$) to $1.85$ ($1.9$) and then reaches $2.05$ ($2.06$). Comparing with Fig.
4, the light curves in Fig. 5 are more diverse at early times because the energy per solid angle is
not a constant. Larger $\theta_0$ will result in a smaller value of energy per solid angle and
cause a lower level of flux density of both the optical and radio afterglows. However, the late
time light curves are almost the same.

We also investigate the effects of $\xi_{\rm{e}}$ and $\xi_{\rm{B}}$ on the light curves. Their
most important effect is in the jet dynamics due to radiative loss, as illustrated in Eqs. (33) and
(34). Figure 6 shows the effects of $\xi_{\rm{e}}$ on the afterglows. In the case of
$\xi_{\rm{e}}=0.2$, $\beta$ changes from $1.2$ to $1.95$. In the case of $\xi_{\rm{e}}=0.4$,
$\beta$ changes from $1.5$ to $2.2$. Both cases show sharp breaks of $\Delta\beta\approx 0.7$ at
around one day. The effect of $\xi_{\rm{B}}$ is more complicated, since changing $\xi_{\rm{B}}$
significantly alters the type of the optical light curve. According to Eq. (32), $t_{\rm{c}}$ is
very sensitive to $\xi_{\rm{B}}$. A small $\xi_{\rm{B}}$ will result in a type B optical afterglow,
in which $\beta=(3p-1)/4$ (Chevalier $\&$ Li 2000). The reason is that the three characteristic
time scales depend on $\xi_{\rm{B}}$ as $t_{\rm{c}}\propto \xi_{\rm{B}}^3$,
$t_{\rm{0}}\propto\xi_{\rm{B}}$ and $t_{\rm{m}}\propto\xi_{\rm{B}}^{1/3}$. Decreasing
$\xi_{\rm{B}}$ will alter these three times significantly and lead to a type B afterglow, i.e.
$t_{\rm{c}}<t_{\rm{0}}<t_{\rm{m}}$. Figure 7 shows that, for $\xi_{\rm{B}}=10^{-2}$ and $10^{-3}$,
the $\beta$ of the optical light curve evolves from $1.6$ at early times to $\sim2.1$ ten days
later. Although $\Delta\beta\approx 0.5$, the break is still obvious since it is completed within
one decade of time in these cases.

It is interesting to study the effect of the electron energy index, $p$, on the light curves.
Theoretically $p$ is very likely to lie within $2.2$ - $2.4$, but observations of GRB afterglows
indicate that $p$ may cover a rather wide range. Figure 8 shows the behavior of the afterglows with
different $p$. It should be pointed out that the jet dynamics is not affected by $p$ in our
considerations. The light curves show no difference at very early times. For the optical afterglow,
$\beta$ evolves from $1.28$ to $2.0$ and $2.25$ at very late times in the case of $p=2.4$, while
from $1.48$ ($1.6$) to $2.2$ ($2.4$) and finally to $2.55$ ($2.8$) in the case of $p=2.6$
($p=2.8$). The former breaks with $\Delta\beta$ $\sim 0.7$ - $0.8$ are sharp enough to make the jet
+ wind model capable of fitting most of the observed breaks in afterglows.

It is also important to determine the jet initial Lorentz factor through fitting the afterglows.
Previous works have discussed the optical flash arising from a reverse shock when a fireball shell
interacts with its circum-progenitor wind (Chevalier $\&$ Li 2000; Wu et al. 2003). The Lorentz
factor can be determined from the optical flash. The imprint of Lorentz factor is also expected to
appear in very early afterglows. Figure 9 illustrates the difference in the early optical and radio
afterglows caused by different initial Lorentz factors. We see that lower Lorentz factors give
lower flux densities at early times. However, late time afterglows depend mainly on the total
energy of the jet and so the light curves in Figure 9 differ from each other only slightly at late
stages.

Lastly, we examine the effects of the viewing angle within our jet + wind model and illustrate the
results in Fig. 10. This study may be of some help to our understanding of orphan afterglows, which
are afterglows whose parent bursts are not observed because they lay off-axis with respect to our
line of sight (Huang, Dai $\&$ Lu 2002).

The actual profile of GRB jets is likely to be Gaussian (Zhang $\&$ M\'{e}sz\'{a}ros 2002a). It is
interesting that the afterglow behavior of a Gaussian jet is very similar to that of a simple
uniform jet, especially when the uniform jet is assumed to have no lateral expansion. This
conclusion is supported by the hydrodynamical simulations of Kumar $\&$ Granot (2003), who showed
that the lateral expansion velocity ($v_{\theta}$) of a Gaussian jet is significantly less than the
local sound speed. In other words, we can approximate a Gaussian jet with a uniform jet with
lateral expansion velocity set at $v_{\theta}=0$. In Fig. 11, we illustrate the effect of
$v_{\theta}$ on the afterglow light curves. It is clearly shown that although the absolute
intensities are different, the jet breaks in the $R$-band light curves are similar in the both
cases. In the $v_{\theta}=0$ case, $\beta$ evolves from $1.0$ to $1.7$, so that the light curve
breaks by about $\Delta\beta=0.7$ at around one day.

\section{DISCUSSION AND CONCLUSIONS}
\label{con} The recently observed association of GRB $030329$ and SN 2003dh strongly suggests a
massive star origin for long GRBs. In the context of massive star evolution, a type Ib/c SN is
expected to explode from a Wolf-Rayet star. The environment is thus likely to be a high speed
stellar wind ejected from the progenitor. Despite the above basic physical reasoning, results of
previous works favored instead an interstellar medium environment. This preference was partly due
to the fact that the previous authors thought that the jet+wind model was unable to produce sharp
breaks in the optical afterglow light curves. In this paper we have revisited a realistic jet+wind
model and studied the effects of various parameters, such as $E_{\rm{iso}}$, $\theta_0$,
$A_{\ast}$, $\xi_{\rm{e}}$, $\xi_{\rm{B}}$, etc.. Our more realistic model includes radiative
energy loss and lateral expansion as well as the equal-arrival time surface effect. Inverse Compton
scattering of synchrotron photons is not considered in this work, which can additionally decrease
the cooling electron Lorentz factor $\gamma_{\rm{c}}$, leading to a further decrease of the jet
energy. We find that obvious breaks can in fact be seen in the optical light curves. Temporal
evolution of the jet energy due to radiative loss may be the main ingredient that gives rise to the
notable break. Inverse Compton spectra components emerge always in the X-ray band, and will not
significantly affect the temporal behavior of the optical afterglows in our calculations. Our
results strongly suggest that breaks in the light curve can also be produced by jets expanding into
stellar winds. The change of temporal index $\Delta\beta$ of these breaks, generally ranging from
$\sim0.6$ to $0.8$, happens from less than a day to several days since the GRB while the
transitions are generally well within one decade of time. The smallest break
$\Delta\beta\approx0.5$ in our calculation occurs in a very weak wind. In reality, the stellar wind
may be surrounded by an outer homogeneous medium, which would greatly complicate the physics of
afterglows.

Despite the overwhelming success of this simple jet model, the structure of GRB jets has recently
been intrigued to the model constructors, although it has already been considered in early works
(M\'{e}sz\'{a}ros et al. 1998; Dai $\&$ Gou 2001). The motivation is driven by the large dispersion
of both the isotropic gamma-ray energy releases and the jet apertures, compared to their tight
correlation resulting in a standard candle as discussed above. There are two simple treatments of
the jet structure that keep to the axial symmetry. One common treatment is to assume the energy per
solid angle and the Lorentz factor (therefore the baryon loading) decrease as power law functions
of the angle from the jet axis. There exists a uniform and very narrow inner cone at the center in
order to keep the total energy not finite. This kind of structured jet has the advantage of being
able to explain the observed larger jet aperture with lower isotropic luminosity simply as the
viewing angle effect, if the power law index of the energy distribution is about $-2$. It is also
capable of reproducing the sharp breaks of some optical afterglows, as well as the observed
luminosity function (Rossi, Lazzati $\&$ Rees 2002; Zhang $\&$ M\'{e}sz\'{a}ros 2002a). As the
transverse gradient of energy density is much smaller in the structured jet than in the homogeneous
jet which has a definite boundary with the environment, the lateral expansion almost never
approaches the local sound speed in the co-moving frame and can essentially be neglected in the
analytic solutions (Kumar $\&$ Granot 2003). Detailed reexamination of structured jet is now
available both numerically (Kumar $\&$ Granot 2003) and analytically (Wei $\&$ Jin 2003; Granot
$\&$ Kumar 2003; Panaitescu $\&$ Kumar 2003; Salmonson 2003). However, the most probable profile of
power law structured jet is the one with the energy index $-2$ and a constant Lorentz factor,
constrained by the existing afterglows (Granot $\&$ Kumar 2003). The other treatment is to assume
the energy per solid angle as a Gaussian function of the angle from the jet axis (Zhang $\&$
M\'{e}sz\'{a}ros 2002a; Zhang et al. 2004). This Gaussian jet, despite relatively small lateral
expansion, will not deviate from the simple homogeneous jet significantly in their afterglow
behaviors (Kumar $\&$ Granot 2003; Zhang $\&$ M\'{e}sz\'{a}ros 2002a).

Although the power law structured jet model has the potential virtue of interpreting the large
dispersion of GRBs within a unified picture, it is still confronted by several difficulties. (1) It
is not supported by the simulations of relativistic jet formation during the core collapsing of
massive stars by MacFadyen $\&$ Woosley (2001). In their J32 model, the profile of the energy
distribution at the emergence from the progenitor envelope is much better described by a Gaussian
function, i.e. a Gaussian jet. In fact, the index should be $-4$ rather than $-2$ between
$5^{\circ}$ and $10^{\circ}$ if a power law energy distribution is assumed. Zhang, Woosley $\&$
Heger (2004) have calculated the propagation of a relativistic jet within its massive stellar
progenitor (see also Zhang, Woosley $\&$ MacFadyen 2003). The ultimate structure at emergence is
characterized by a uniform core component of a higher Lorentz factor spanning a few degree, with a
sharp decline boundary adjoining a wider component with a lower Lorentz factor. (2) The power law
structured jet model unifies most GRBs at the expense of introducing another adjustable parameter,
$\theta_{\rm{c}}$, the half-opening angle of the uniform core component. Different central engine
properties will lead to different jet half-opening angle $\theta$ (MacFadyen $\&$ Woosley 2001). To
maintain the merit of power law structured jet model we have to regard $\theta_{\rm{c}}$ as a
universal constant for all GRBs, and this has to be confirmed by observations through data fitting.
(3) The most probable distribution of the Lorentz factor is a homogeneous one within the jet. But
it is difficult to imagine such a distribution can be sustained at large angles and not affected by
abundant baryon loading outside a very narrow funnel. The homogeneous jet model also appreciably
suffers from this problem for a few GRB afterglows with large apertures, and can be settled by
considering a two-component model as recently proposed (Berger et al 2003b; Huang, et al. 2004).
(4) Another difficulty for the power law structured jet model may be the pre-break flattening
predicted by a few authors (Wei $\&$ Jin 2003; Kumar $\&$ Granot 2003; Granot $\&$ Kumar 2003). The
flattening is due to the emergence of the core component before the jet break of the light curve
while the observer's line of sight is outside $\theta_{\rm{c}}$. Such a flattening will even
develop to a pre-break bump if $\theta_{\rm{c}}$ is within $1^{\circ}$ (Salmonson 2003). This new
type of bump has been proposed to explain the peculiar behavior of GRB 000301C, as an alternative
to energy injection, density jump and microlensing scenarios (Dai $\&$ Lu 1998a; Zhang $\&$
M\'{e}sz\'{a}ros 2002b; Dai $\&$ Lu 2002; Garnavich, Loeb $\&$ Stanek 2000). However, it is suspect
because some well-observed afterglows with large jet aperture such as GRB $000926$
($\theta_{\rm{j}}=8.1^{\circ}$) have not shown this pre-break bump behavior (Panaitescu $\&$ Kumar
2002). Future observations of very early afterglows in the upcoming $Swift$ era would help to
discriminate the structure of GRB jets. As proposed by Zhang $\&$ M\'{e}sz\'{a}ros (2002a), actual
profiles of GRB jets may be Gaussian. The sideways expansion can be neglected since the gradients
of physical parameters in a Gaussian jet are much smaller than in an ideal uniform jet with a
clear-cut lateral boundary to its environment. In such a case, a sharp break exists as well. With
the self-absorption of synchrotron radiation included in our realistic jet+wind model, we are able
to do broad-band fittings of observed GRB afterglows.

\begin{acknowledgements}
We thank the anonymous referee for useful comments. This work was supported by the National Natural
Science Foundation of China (grant numbers 10233010, 10003001, 10221001), the National 973 project
(NKBRSF G19990754), the Foundation for the Authors of National Excellent Doctorial Dissertations of
P. R. China (Project No: 200125) and the Special Funds for Major State Basic Research Projects.
\end{acknowledgements}

%-------------------------------------------------------------------

%--------------------------------------------------%figure
\begin{figure}
  \includegraphics[width=130mm,height=100mm]{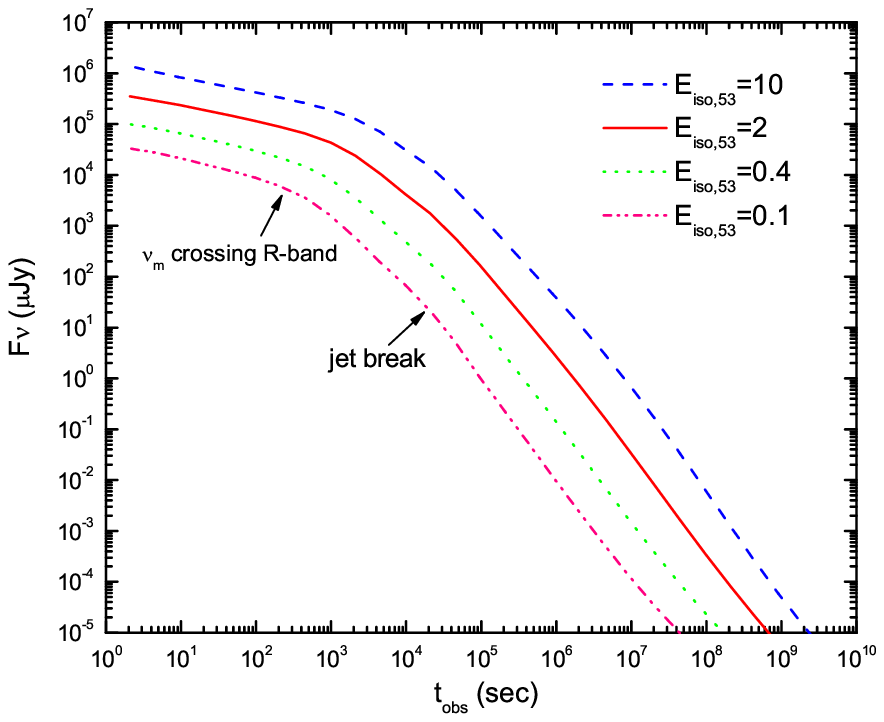}
  \includegraphics[width=130mm,height=100mm]{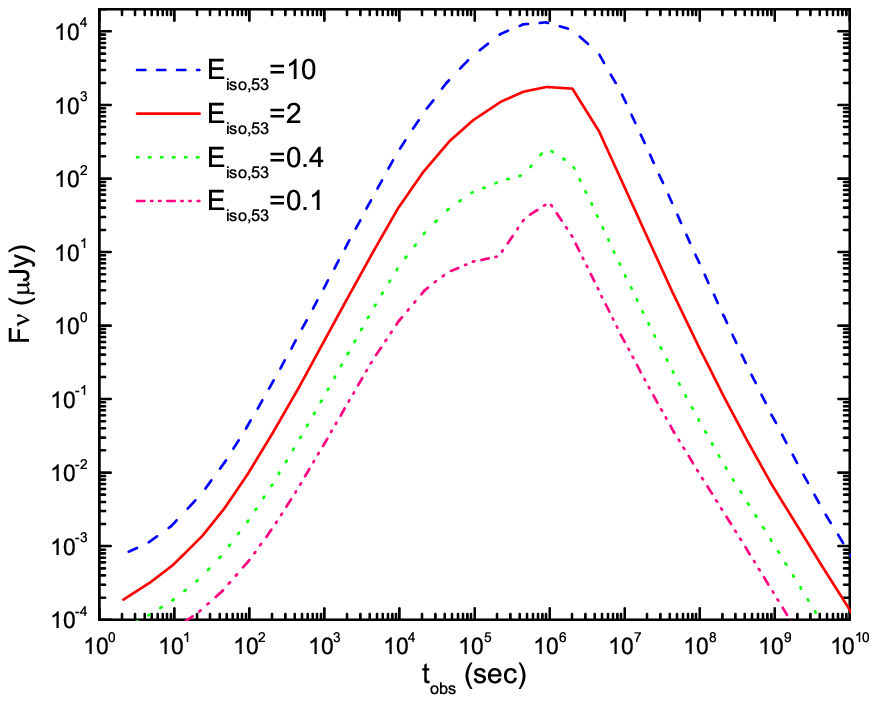}
  \caption{Effects of the parameter $E_{\rm{iso}}$ on the optical ($R-$band, upper panel) and radio ($4.86$ GHz,
           lower panel) light curves. The solid line corresponds to a `standard' jet with $E_{\rm{iso}}=2
           \times 10^{53}$ erg, $\theta_0=0.1$, $\gamma_0=300$, $\xi_{\rm{e}}=0.1$, $\xi_{\rm{B}}=0.1$, and
           $p=2.2$ running into a stellar wind with $A_{\ast}=1.0$. Other lines are drawn with only $E_{\rm{iso}}$
           changed.}
  \label{Fig1}
\end{figure}
%--------------------------------------------------%figure 1
\begin{figure}
  \includegraphics[width=130mm,height=100mm]{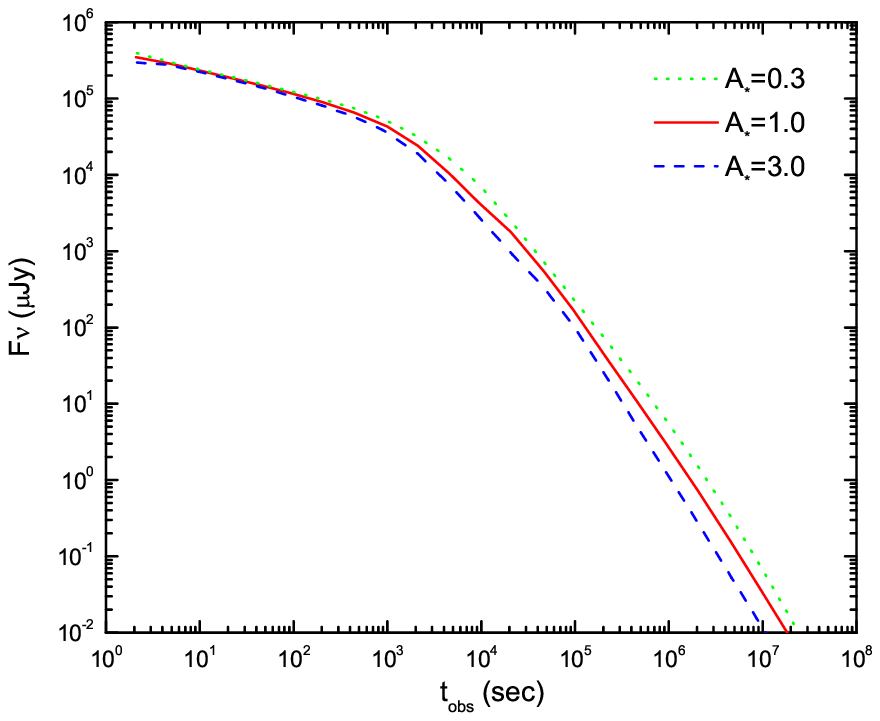}
  \includegraphics[width=130mm,height=100mm]{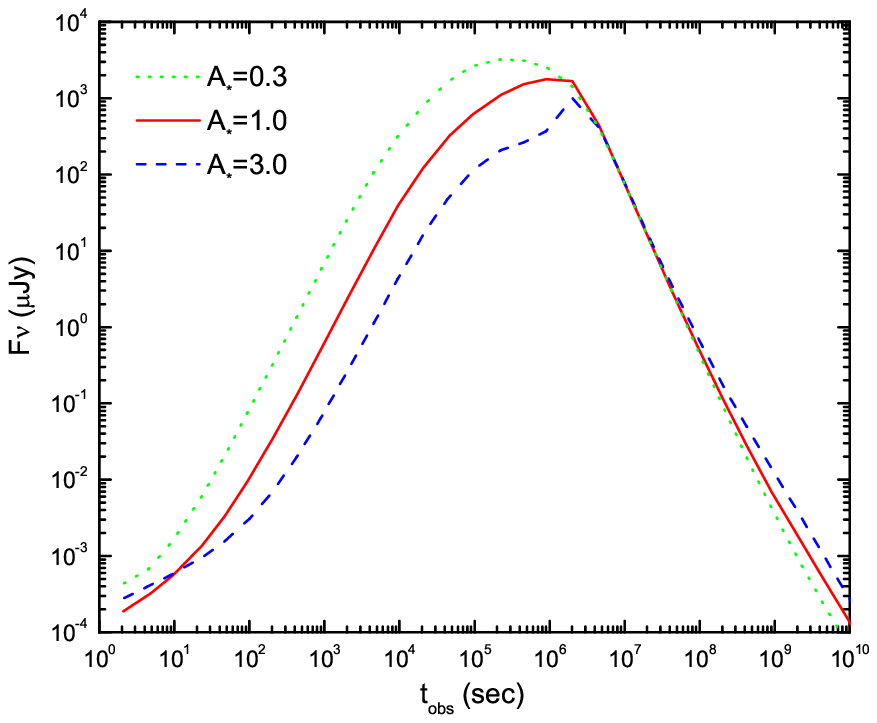}
  \caption{Effect of the parameter $A_{\ast}$ on the optical ($R-$band, upper panel) and radio ($4.86$ GHz, lower panel)
           light curves. The solid line corresponds to a `standard' jet with $E_{\rm{iso}}=2\times 10^{53}$ erg,
           $\theta_0=0.1$, $\gamma_0=300$, $\xi_{\rm{e}}=0.1$, $\xi_{\rm{B}}=0.1$, and $p=2.2$
           running into a stellar wind with $A_{\ast}=1.0$. Other lines are drawn with only $A_{\ast}$ changed.}
  \label{Fig2}
\end{figure}
%--------------------------------------------------%figure 2
\begin{figure}
  \includegraphics[width=130mm,height=100mm]{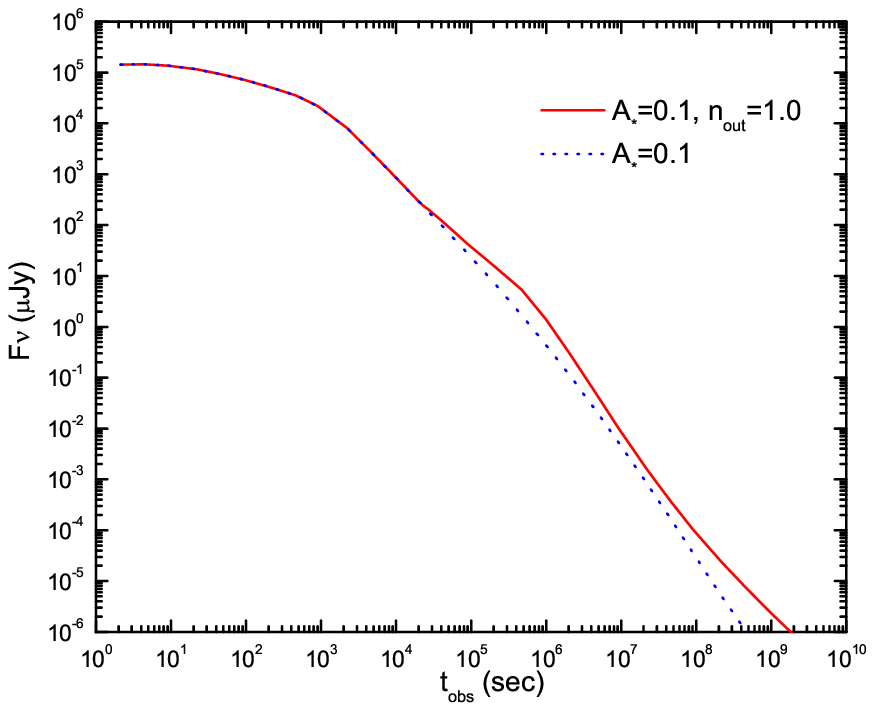}
  \includegraphics[width=130mm,height=100mm]{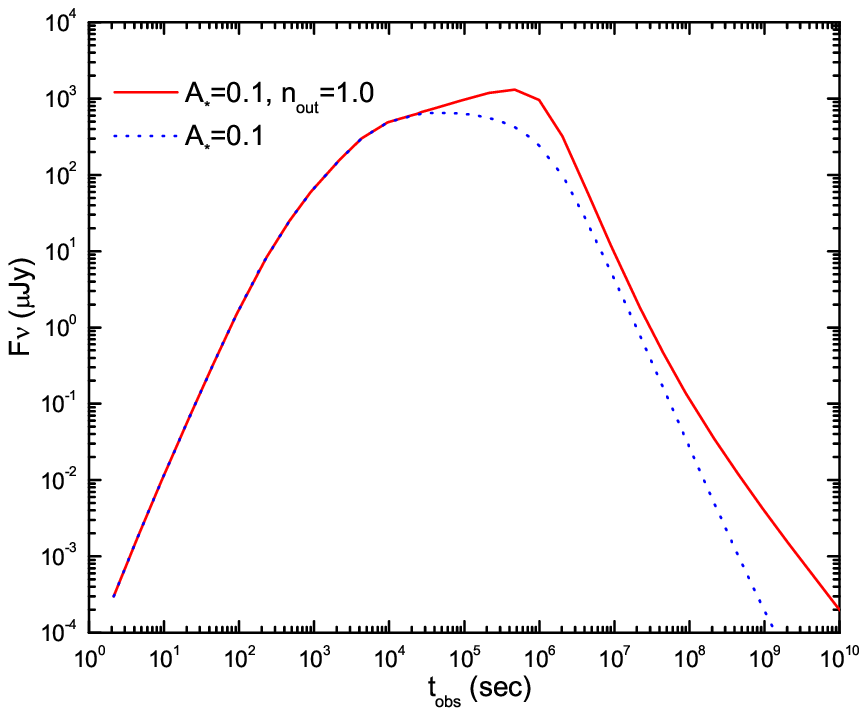}
  \caption{Optical ($R-$band, upper panel) and radio ($4.86$ GHz, lower panel) afterglows from jets in a wind environment
           surrounded by a uniform medium. The solid line corresponds to a jet with $E_{\rm{iso}}=2\times 10^{53}$ erg,
           $\theta_0=0.1$, $\gamma_0=300$, $\xi_{\rm{e}}=0.1$, $\xi_{\rm{B}}=0.1$, and $p=2.2$ running into a
           circum-stellar wind with $A_{\ast}=0.1$ and then entering the outer uniform medium of $n_{\rm{out}}=1.0$
           cm$^{-3}$ when $R>R_{\rm{c}}=4\times 10^{17}$ cm. An entire stellar wind situation is calculated (dotted line)
           for comparison.}
  \label{Fig3}
\end{figure}
%--------------------------------------------------%figure 3
\begin{figure}
  \includegraphics[width=130mm,height=100mm]{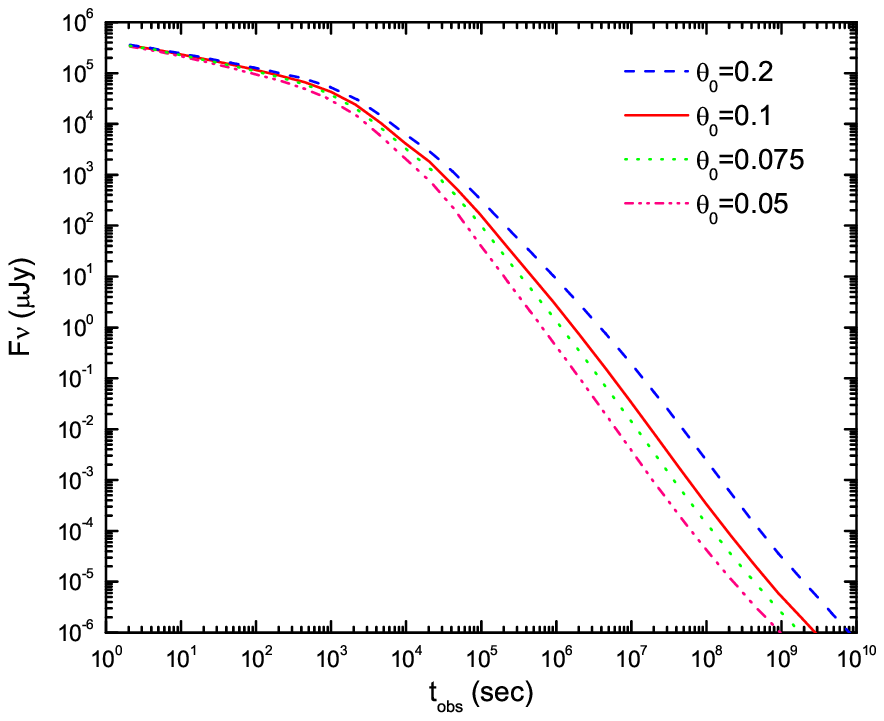}
  \includegraphics[width=130mm,height=100mm]{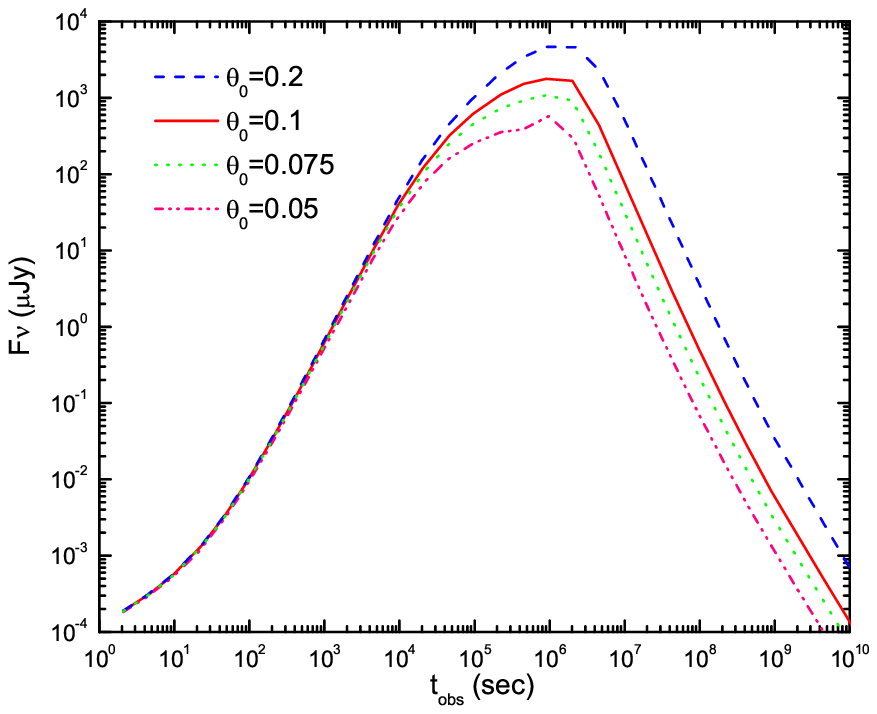}
  \caption{Effects of the parameter $\theta_0$ on the optical ($R-$band, upper panel) and radio ($4.86$ GHz, lower panel)
           light curves. The solid line corresponds to a `standard' jet with $E_{\rm{iso}}=2\times 10^{53}$ erg,
           $\theta_0=0.1$, $\gamma_0=300$, $\xi_{\rm{e}}=0.1$, $\xi_{\rm{B}}=0.1$, and $p=2.2$
           running into a stellar wind with $A_{\ast}=1.0$. Other lines are drawn with only $\theta_0$ changed.}
  \label{Fig4}
\end{figure}
%--------------------------------------------------%figure 4
\begin{figure}
  \includegraphics[width=130mm,height=100mm]{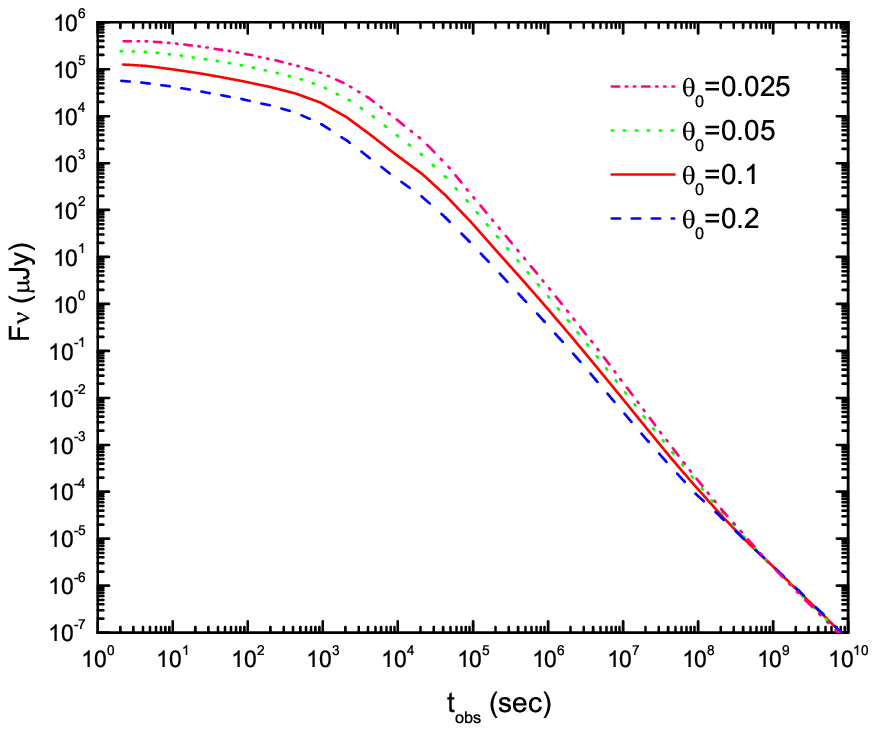}
  \includegraphics[width=130mm,height=100mm]{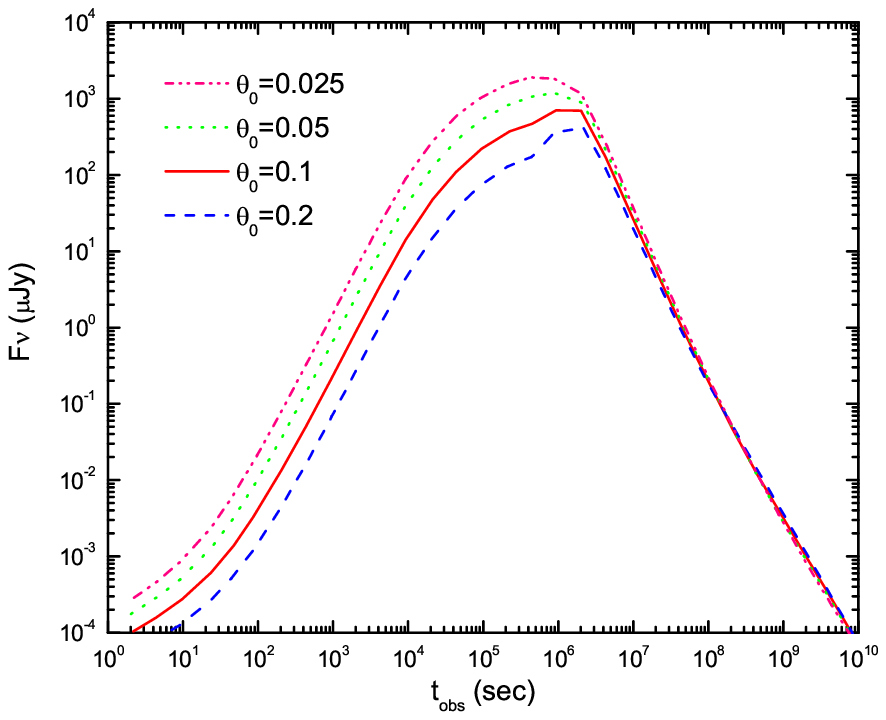}
  \caption{Optical ($R-$band, upper panel) and radio ($4.86$ GHz, lower panel) afterglows from jets under the standard
           energy reservoir assumption, i.e. $E_{\rm{j}}=E_{\rm{iso}}\displaystyle\frac{\theta_0^2}{2}\equiv2\times10^{51}$
           erg. The solid line corresponds to a `standard' jet with $E_{\rm{iso}}=2\times 10^{53}$ erg,
           $\theta_0=0.1$, $\gamma_0=300$, $\xi_{\rm{e}}=0.1$, $\xi_{\rm{B}}=0.1$, and $p=2.2$
           running into a stellar wind with $A_{\ast}=1.0$. Other lines are drawn with only $\theta_{0}$ changed while
           keeping $E_{\rm{j}}$ the same.}
  \label{Fig5}
\end{figure}
%--------------------------------------------------%figure 5
\begin{figure}
  \includegraphics[width=130mm,height=100mm]{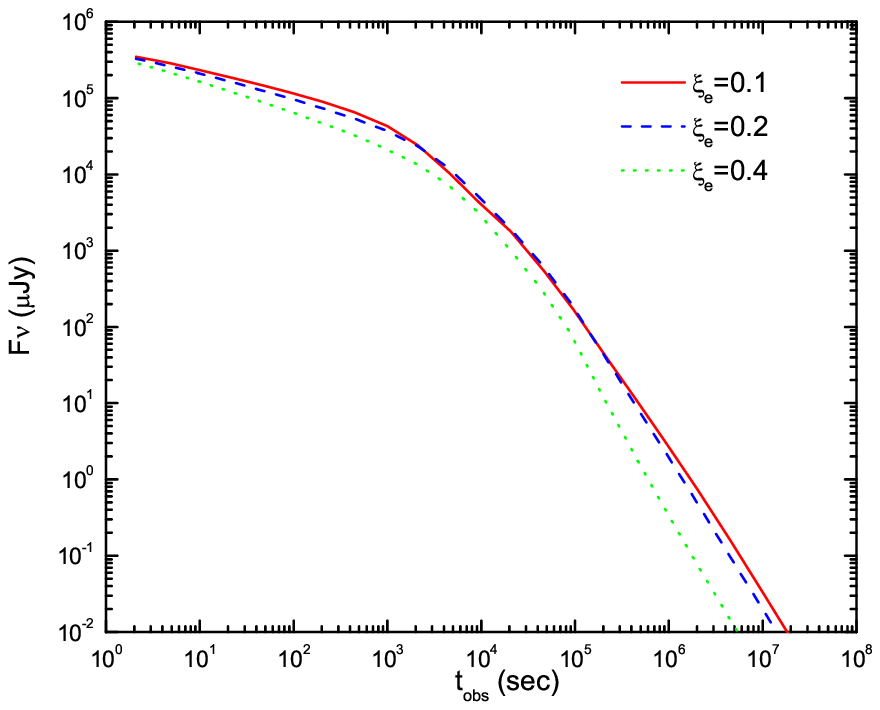}
  \includegraphics[width=130mm,height=100mm]{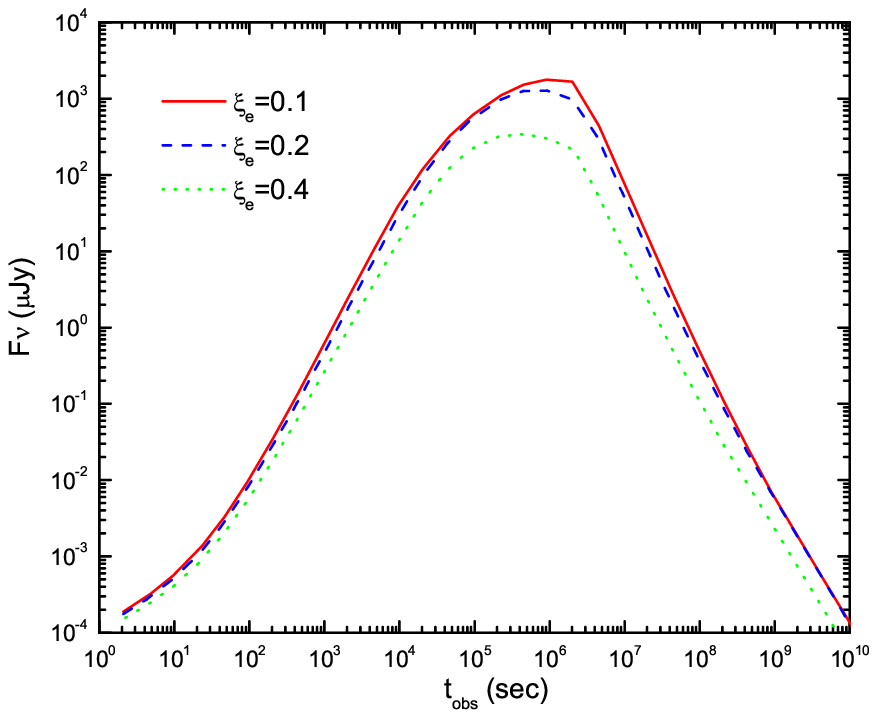}
  \caption{Effects of the parameter $\xi_{\rm{e}}$ on the optical ($R-$band, upper panel) and radio ($4.86$ GHz, lower panel)
           light curves. The solid line corresponds to a `standard' jet with $E_{\rm{iso}}=2\times 10^{53}$ erg,
           $\theta_0=0.1$, $\gamma_0=300$, $\xi_{\rm{e}}=0.1$, $\xi_{\rm{B}}=0.1$, and $p=2.2$ running into a stellar
           wind with $A_{\ast}=1.0$. Other lines are drawn with only $\xi_{\rm{e}}$ changed.}
  \label{Fig6}
\end{figure}
%--------------------------------------------------%figure 6
\begin{figure}
  \includegraphics[width=130mm,height=100mm]{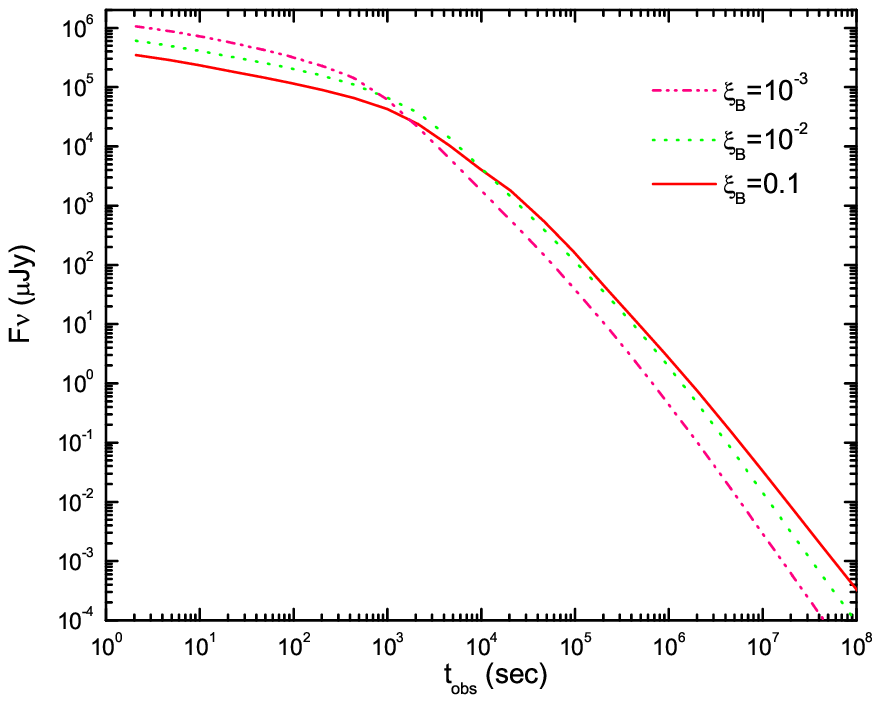}
  \includegraphics[width=130mm,height=100mm]{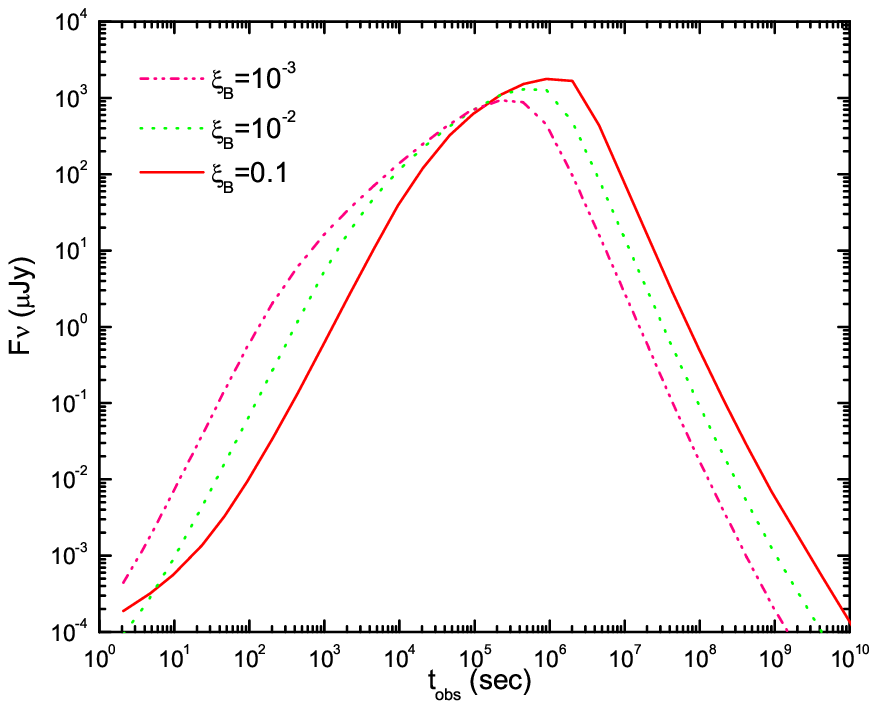}
  \caption{Effects of the parameter $\xi_{\rm{B}}$ on the optical ($R-$band, upper panel) and radio ($4.86$ GHz, lower panel)
           light curves. The solid line corresponds to a `standard' jet with $E_{\rm{iso}}=2\times 10^{53}$ erg,
           $\theta_0=0.1$, $\gamma_0=300$, $\xi_{\rm{e}}=0.1$, $\xi_{\rm{B}}=0.1$, and $p=2.2$ running into a stellar
           wind with $A_{\ast}=1.0$. Other lines are drawn with only $\xi_{\rm{B}}$ changed.}
  \label{Fig7}
\end{figure}
%--------------------------------------------------%figure 7
\begin{figure}
  \includegraphics[width=130mm,height=100mm]{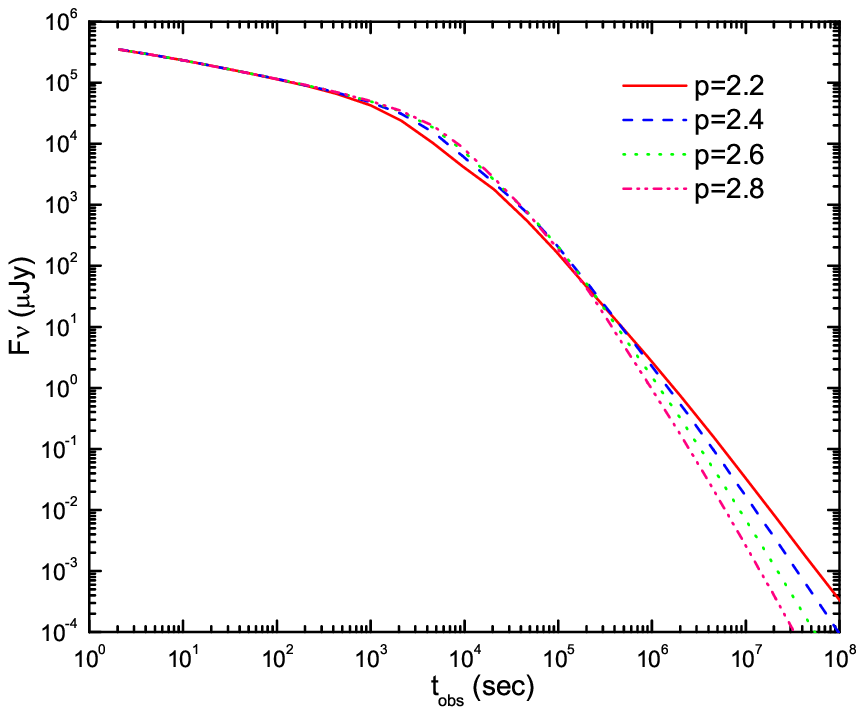}
  \includegraphics[width=130mm,height=100mm]{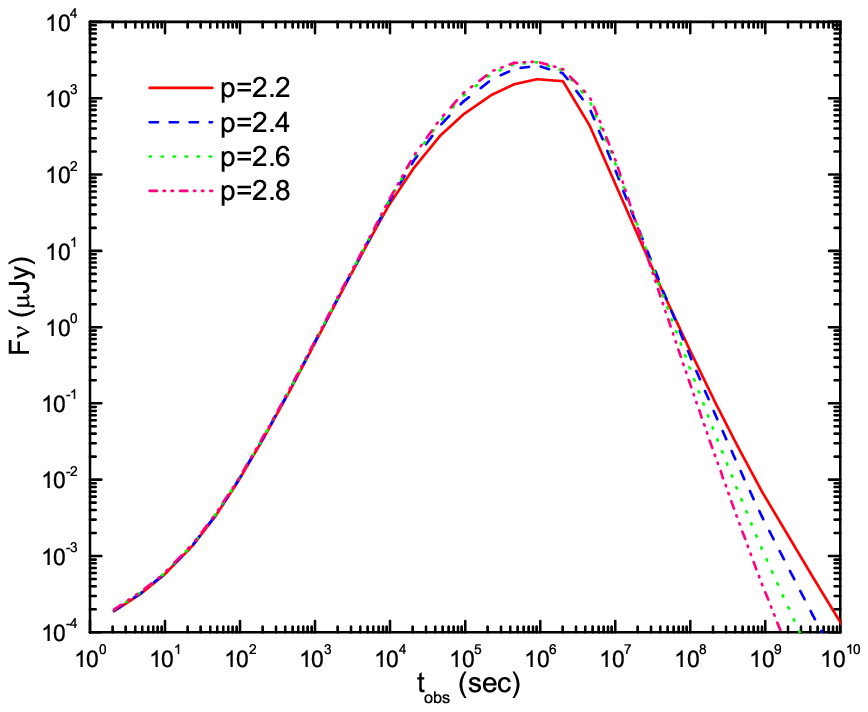}
  \caption{Effects of the parameter $p$ on the optical ($R-$band, upper panel) and radio ($4.86$ GHz, lower panel)
           light curves. The solid line corresponds to a `standard' jet with $E_{\rm{iso}}=2\times 10^{53}$ erg,
           $\theta_0=0.1$, $\gamma_0=300$, $\xi_{\rm{e}}=0.1$, $\xi_{\rm{B}}=0.1$, and $p=2.2$ running into a stellar
           wind with $A_{\ast}=1.0$. Other lines are drawn with only $p$ changed.}
  \label{Fig8}
\end{figure}
%--------------------------------------------------%figure 8
\begin{figure}
  \includegraphics[width=130mm,height=100mm]{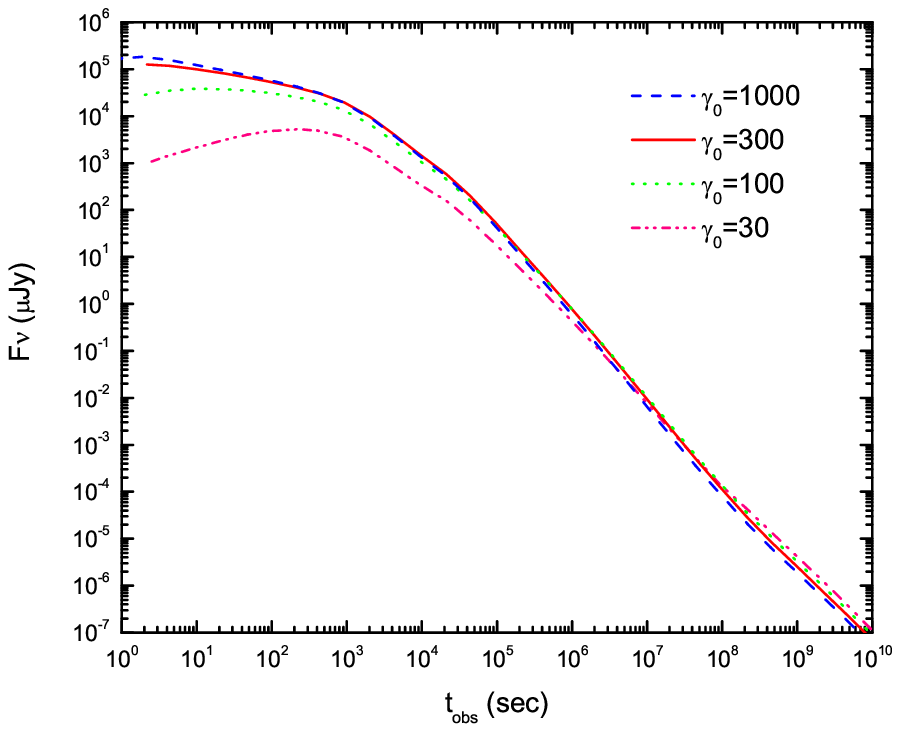}
  \includegraphics[width=130mm,height=100mm]{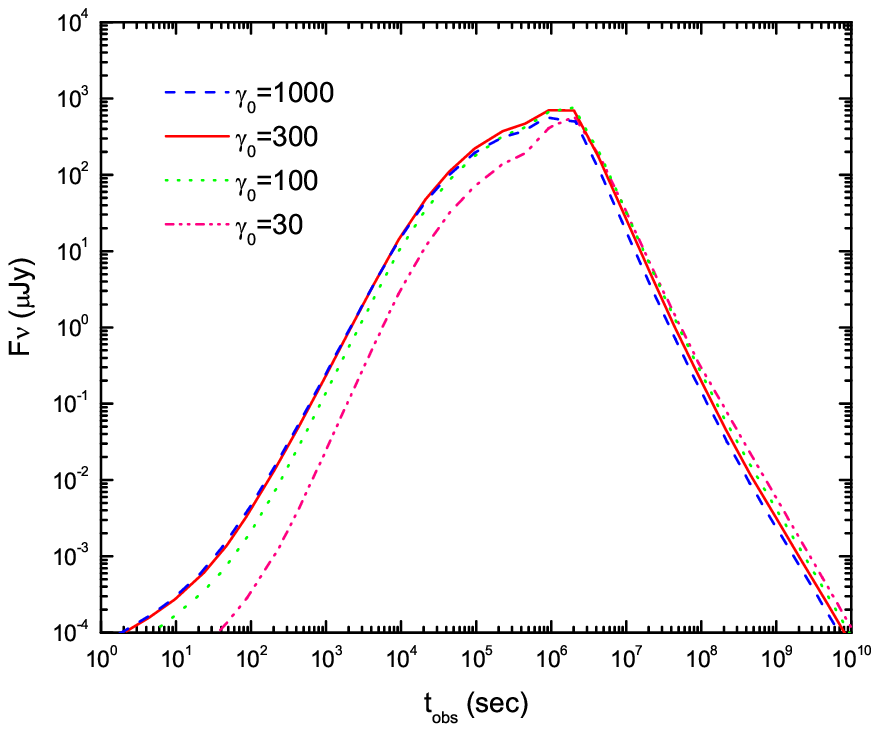}
  \caption{Effects of the parameter $\gamma_0$ on the optical ($R-$band, upper panel) and radio ($4.86$ GHz, lower panel)
           light curves. The solid line corresponds to a `standard' jet with $E_{\rm{iso}}=2\times 10^{53}$ erg,
           $\theta_0=0.1$, $\gamma_0=300$, $\xi_{\rm{e}}=0.1$, $\xi_{\rm{B}}=0.1$, and $p=2.2$ running into a stellar
           wind with $A_{\ast}=1.0$. Other lines are drawn with only $\gamma_0$ changed.}
  \label{Fig9}
\end{figure}
%--------------------------------------------------%figure 9
\begin{figure}
  \includegraphics[width=130mm,height=100mm]{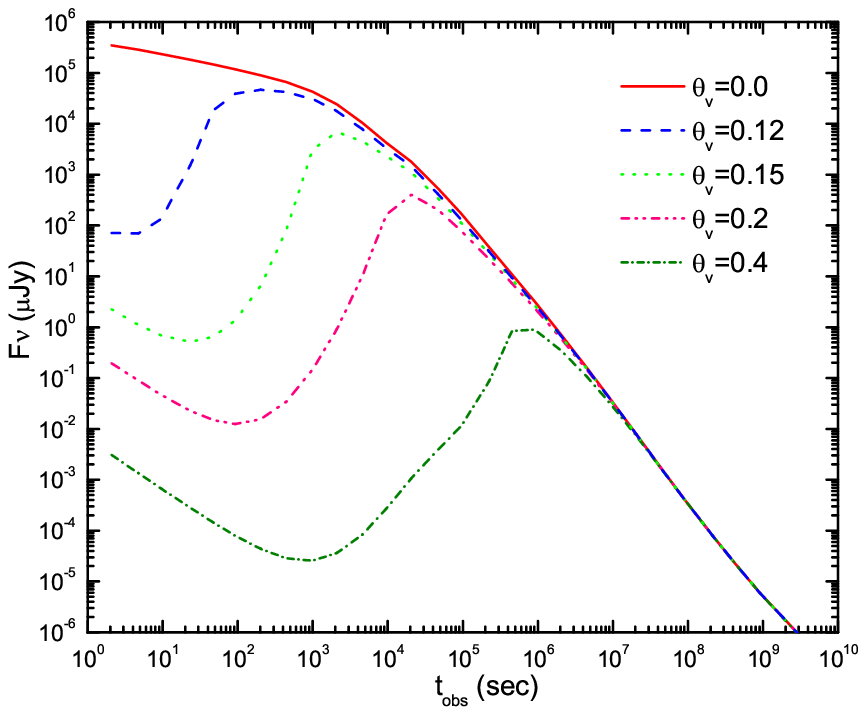}
  \includegraphics[width=130mm,height=100mm]{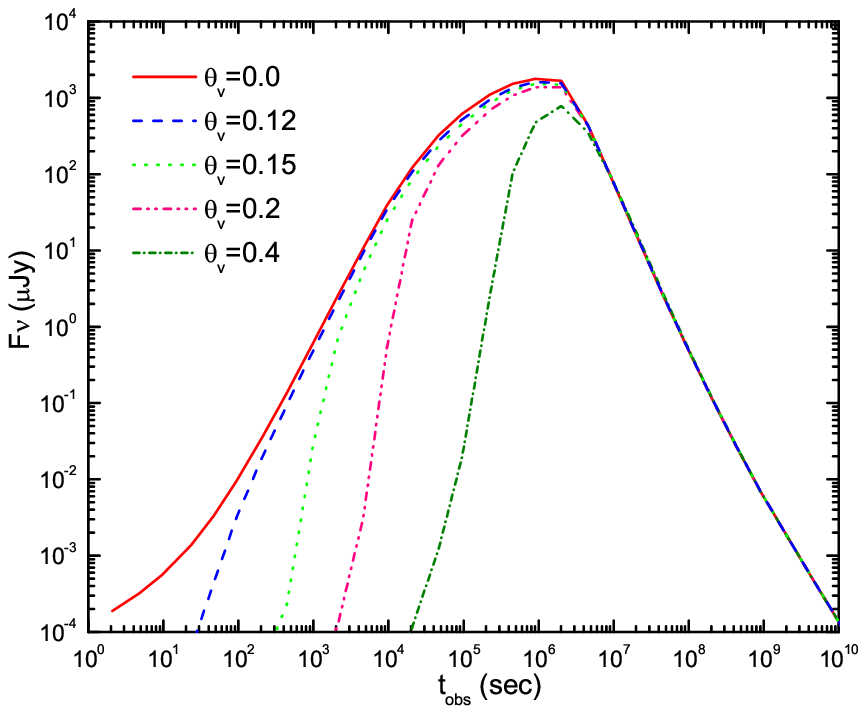}
  \caption{Effects of the parameter $\theta_{\rm{v}}$ on the optical ($R-$band, upper panel) and radio ($4.86$ GHz, lower panel)
           light curves. The solid line corresponds to a `standard' jet with $E_{\rm{iso}}=2\times 10^{53}$ erg,
           $\theta_0=0.1$, $\gamma_0=300$, $\xi_{\rm{e}}=0.1$, $\xi_{\rm{B}}=0.1$, and $p=2.2$ running into a stellar
           wind with $A_{\ast}=1.0$. Other lines are drawn with only $\theta_{\rm{v}}$ changed.}
  \label{Fig10}
\end{figure}
%--------------------------------------------------%figure 10
\begin{figure}
  \includegraphics[width=130mm,height=100mm]{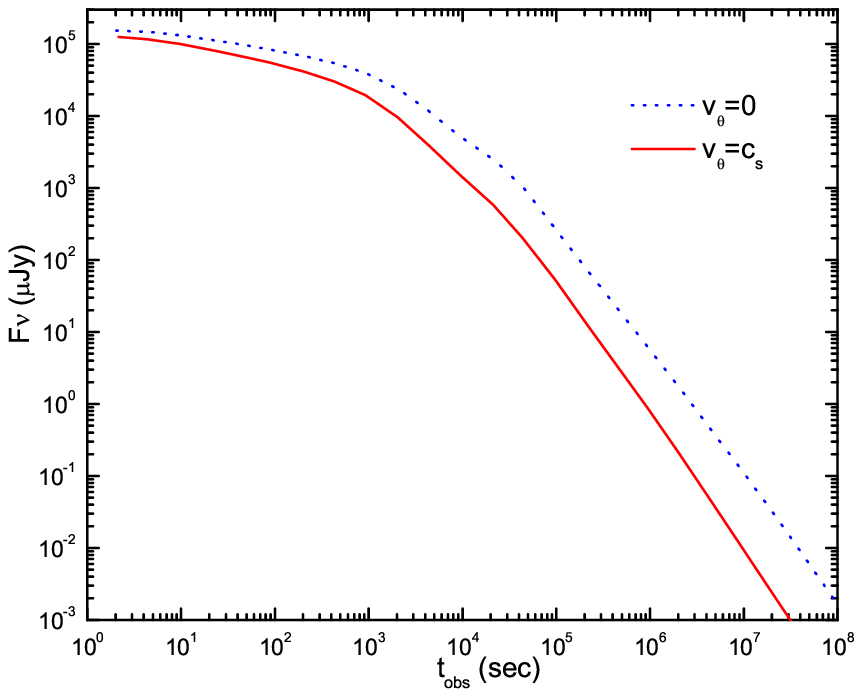}
  \includegraphics[width=130mm,height=100mm]{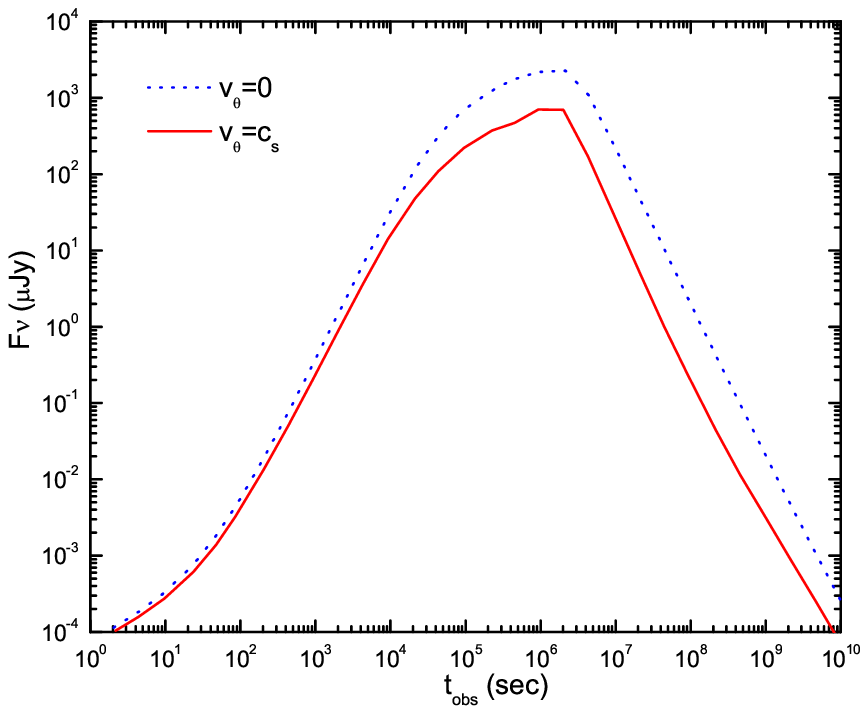}
  \caption{Effects of lateral expansion velocity $v_{\theta}$ on the optical ($R-$band, upper panel) and
           radio ($4.86$ GHz, lower panel) light curves. The solid line corresponds to a `standard' jet
           with $E_{\rm{iso}}=2\times 10^{53}$ erg, $\theta_0=0.1$, $\gamma_0=300$, $\xi_{\rm{e}}=0.1$,
           $\xi_{\rm{B}}=0.1$, and $p=2.2$ running into a stellar wind with $A_{\ast}=1.0$, with $v_{\theta}=c_{\rm{s}}$.
           The dotted line corresponds to no lateral expansion, i.e. $v_{\theta}=0$.}
  \label{Fig11}
\end{figure}
%--------------------------------------------------%figure 11

\end{document}